\documentclass[apjl]{emulateapj}
\usepackage{amsfonts,amsmath,graphicx,natbib,apjfonts,subfigure,color,gensymb,longtable}

\def\gw{GW\,170817 }
\def\nuu{2}
\def\cfa{6}
\def\hf{5}
\def\fr{1}

\def\oh{7}
\def\fermi{8}
\def\penn{3}
\def\kavli{9}

\def\kavlidan{4}
\def\brandeis{11}
\def\Uchi{9}

\def\mich{10}

\def\hinvGpc{\,h^{-1}{\rm Gpc}}
\def\Msun{M_\odot}
\def\kmsMpc{\,{\rm km/s/Mpc}}
\def\hinvMsun{\,h^{-1}{\rm \Msun}}

\newcommand{\hunits}{\mbox{km/s/Mpc}}
\newcommand{\incangle}{\degree}

\newcommand{\finalhabbott}{$70.0 \pm \frac{12.0}{8.0}~$}

\newcommand{\finalhbaseline}{$75.5 \pm \frac{11.6}{9.6}$}
\newcommand{\finalhligohi}{$70.5 \pm \frac{12.8}{9.6}$}
\newcommand{\finalhligolow}{$69.4 \pm \frac{12.0}{7.7}$}
\newcommand{\finalhfive}{$72.0 \pm \frac{7.6}{8.0}$}

\newcommand{\finalhthetalim}{$70.5 \pm \frac{12.7}{8.9}$}
\newcommand{\finalhbaselow}{$74.0 \pm \frac{11.5}{7.5}$}
\newcommand{\finalhfivelow}{$71.0 \pm \frac{7.1}{5.7}$}
\shorttitle{\gw and $H_0$ }
\shortauthors{Guidorzi~et~al.}

\begin{document}
\title{Improved Constraints on H0 from a combined analysis of gravitational-wave and electromagnetic emission from GW170817}

\author{C.~Guidorzi\altaffilmark{\fr}, R.~Margutti\altaffilmark{\nuu}, D. Brout\altaffilmark{\penn}, D. Scolnic\altaffilmark{\kavlidan,\hf}, W.~Fong\altaffilmark{\nuu}$^,$\altaffilmark{\hf}, K.~D.~Alexander\altaffilmark{\cfa}, P.~S.~Cowperthwaite\altaffilmark{\cfa}, J.~Annis\altaffilmark{\fermi}, E.~Berger\altaffilmark{\cfa}, P.~K.~Blanchard\altaffilmark{\cfa}, R.~Chornock\altaffilmark{\oh}, D.~L. Coppejans\altaffilmark{\nuu}, T.~Eftekhari\altaffilmark{\cfa}, J.~A.~Frieman\altaffilmark{\fermi,\kavli}, D.~Huterer\altaffilmark{\mich}, M.~Nicholl\altaffilmark{\cfa},  M. Soares-Santos\altaffilmark{\brandeis,\fermi}, G. Terreran\altaffilmark{\nuu}, V.~A.~Villar\altaffilmark{\cfa}, P.~K.~G.~Williams\altaffilmark{\cfa}
}
\altaffiltext{\fr}{Department of Physics and Earth Science, University of Ferrara, via Saragat 1, I--44122, Ferrara, Italy}
\altaffiltext{\nuu}{Center for Interdisciplinary Exploration and Research in Astrophysics (CIERA) and Department of Physics and Astronomy, Northwestern University, Evanston, IL 60208}
\altaffiltext{\penn}{Department of Physics and Astronomy, University of Pennsylvania, Philadelphia, PA 19104, USA}
\altaffiltext{\kavlidan}{Enrico Fermi Institute, Department of Physics, Department of Astronomy and Astrophysics, University of Chicago, Chicago, IL 60637, USA}
\altaffiltext{\hf}{Hubble Fellow}
\altaffiltext{\cfa}{Harvard-Smithsonian Center for Astrophysics, 60 Garden Street, Cambridge, MA 02138, USA}

\altaffiltext{\oh}{Astrophysical Institute, Department of Physics and Astronomy, 251B Clippinger Lab, Ohio University, Athens, OH 45701, USA}
\altaffiltext{\fermi}{Fermi National Accelerator Laboratory, P.O. Box 500, Batavia, IL 60510, USA}
\altaffiltext{\Uchi}{Department of Astronomy and Astrophysics, University of Chicago, Chicago, Illinois 60637, USA}
\altaffiltext{\mich}{Department of Physics, University of Michigan, 450 Church Street, Ann Arbor, MI 48109, USA}
%\altaffiltext{\kavli}{Kavli Institute for Cosmological Physics, University of Chicago, Chicago, IL 60637, USA}
\altaffiltext{\brandeis}{Department of Physics, Brandeis University, Waltham, MA 02454, USA}

\begin{abstract}
The luminosity distance measurement of GW170817 derived from GW analysis in \cite{LIGOH0} (A17:H0) is highly correlated with the measured inclination of the NS-NS system.  To improve the precision of the distance measurement, we attempt to constrain the inclination by modeling the broad-band X-ray-to-radio emission from GW170817, which is dominated by the interaction of the jet with the environment. We update our previous analysis and we consider the radio and X-ray data obtained at $t<40$ days since merger. We find that the afterglow emission from  GW170817 is consistent with an off-axis relativistic jet with energy $10^{48}\,\rm{erg}<E_{k}\le 3\times 10^{50} \,\rm{erg}$ propagating into an environment with density $n\sim10^{-2}-10^{-4} \,\rm{cm^{-3}}$, with preference for wider jets (opening angle $\theta_j=15\degree$).  For these jets, our modeling indicates an off-axis angle $\theta_{\rm obs}\sim25\degree-50\degree$.  We combine our constraints on $\theta_{\rm obs}$ with the joint distance-inclination constraint from LIGO. Using the same $\sim 170$ km/sec peculiar velocity uncertainty assumed in A17:H0 but with an inclination constraint from the afterglow data, we get a value of $H_0=$$74.0 \pm \frac{11.5}{7.5}$ $\mbox{km/s/Mpc}$, which is higher than the value of $H_0=$$70.0 \pm \frac{12.0}{8.0}$ $\mbox{km/s/Mpc}$ found in A17:H0. Further, using a more realistic peculiar velocity uncertainty of 250 km/sec derived from previous work, we find $H_0=$$75.5 \pm \frac{11.6}{9.6}$ km/s/Mpc for H0 from this system. We note that this is in modestly better agreement with the local distance ladder than the Planck CMB, though a significant such discrimination will require $\sim 50$ such events. Future measurements at $t>100$ days of the X-ray and radio emission will lead to tighter constraints. 
\end{abstract}
%%%%%%%%%%%%%%%%%%%%%%%%%%%%%%%%%%%%%%%%%%%
\section{Introduction}
\label{Sec:intro}

The first gravitational wave (GW) detection of a binary neutron star (BNS) merger was made on  2017 August 17 at 12:41:02 UT by the Advanced LIGO/Virgo detectors \citep{ALVgcn,ALVdetection}.  The event was localized to a region of about 30 deg$^2$ with an estimated distance of $\sim40$ Mpc \citep{ALVgcn}. A short burst of $\gamma$-rays was detected with a delay of about 2 s relative to the merger time by {\it Fermi}/GBM and Integral \citep{GBMgcn1,INTEGRALgcn,GBMdetection,INTEGRALdetection}.  Optical follow-up observations of the LIGO/Virgo sky map, led to several independent detections of a counterpart, associated with the galaxy NGC\,4993 \citep{Coulter,ValentiPaper,DECamPaper1,PhilPaper,MattPaper,RyanPaper}.  

It has long been argued that the combination of a distance measurement from a gravitational wave signal with a redshift measurement from an electromagnetic counterpart can be used to measure cosmological parameters in a novel way, in particular the Hubble constant \citep{Schutz,Holz5,Dalal}.  Here, the ratio of the redshift velocity of the host galaxy to the absolute distance from the GW event directly yields the Hubble constant such that $v=H_0 \times d$ where v is in $km/s$, $H_0$ is in \hunits, and $d$ is in Mpc.   This method has been discussed in \cite{Nissanke} and \cite{Taylor12}, which show that with large numbers (20-50) of similar GW-EM detections out to $z\sim0.1$, percent-level $H_0$ measurements can be determined. This measurement is important as there is currently $3.4\sigma$ tension between the local measurement of $H_0$ \citep{riessetal16} and the CMB measurement of $H_0$ \citep{Planck16_cosmo}.  The tension may be a hint of new cosmological physics, so independent measurements of $H_0$ are needed to resolve this issue.

The first BNS event discovered by LIGO-Virgo allows for the first independent measurement of the Hubble constant using GWs (\citealt{LIGOH0}, hereafter A17:H0).  While previous studies are correct in finding that a number of events are needed to make a percent level measurement, the current tension in $H_0$ measurements is large enough that even single GW events could provide interesting constraints.  However, the key limitation in the precision of GW $H_0$ measurement by 
A17:H0 is due to the degeneracy between the distance, which is $\sim40$ Mpc, and the orbital inclination angle of the BNS. In this context, the inclination is defined as the angle between the line of sight from source to earth and the angular momentum of the binary system. 

The analysis of the electromagnetic emission from BNS mergers provides an independent constraint on the inclination angle, as these systems are expected to launch relativistic jets aligned with their angular momentum vector (e.g. \citealt{Eichler89,Narayan92}). The interaction of a relativistic jet with material in the circumbinary environment is a well known  source of non-thermal synchrotron emission across the electromagnetic spectrum (e.g. \citealt{Sari98,Sari99,Sari99b,Granot02}), known as ``afterglow'' in the Gamma-Ray Burst (GRB) literature. While the UV-optical-NIR emission from a BNS merger can be dominated by the ``kilonova'' (i.e. a transient powered by the radioactive decay of r-process nuclei synthesized in the neutron-rich merger ejecta, \citealt{Metzger17}), the jet interaction with the circumbinary medium dominates the X-ray and radio portion of the electromagnetic spectrum, with observable properties that directly depend on the binary inclination angle with respect to the line of sight (e.g. \citealt{Granot02b,Rossi2002}). 

In this paper we build on our previous analysis of the X-ray and radio emission from \gw  (\citealt{MyPaper,KatePaper}; hereafter, M17 and A17, respectively). We update our modeling to include all the available data obtained at $t<40$ days since NS-NS coalescence, and we run a more extended and finer grid of off-axis jets simulations. Our data set is described in Sec. \ref{Sec:data}, while in Sec. \ref{Sec:afterglow} we employ realistic models of synchrotron emission from off-axis relativistic jets to estimate the BNS jet parameters. We combine the estimate of the binary inclination angle obtained with these models with the GW measurement and improve the distance determination of \gw in Sec. \ref{Sec:H0}. We conclude in Sec. \ref{Sec:Conc}. With this pilot study we demonstrate that the combination of GW and EM observations of the \emph{same} BNS merger improves the accuracy of the measurement of cosmological parameters. 

%Briefly, we can leverage the knowledge from this and that to constrain inclination angle to the order of this much.

%For this analysis, we assume a distance to NGC\,4993 of 39.5\,Mpc ($z=0.00973$) as listed in the NASA Extragalactic Database - though the uncertainty on the redshift is discussed in detail below. \textbf{We will have to vary Dl as well. There is some limited dependencies here} 

In this paper we list $1\,\sigma$ c.l. 
uncertainties unless otherwise stated and employ the notation $Q_x\equiv Q/10^x$.

%%%%%%%%%%%%%%%%%%%%%%%%%%%%%%%%%%%%%%%%%%
\begin{figure*}[t!]
\center
 \includegraphics[scale=0.435]{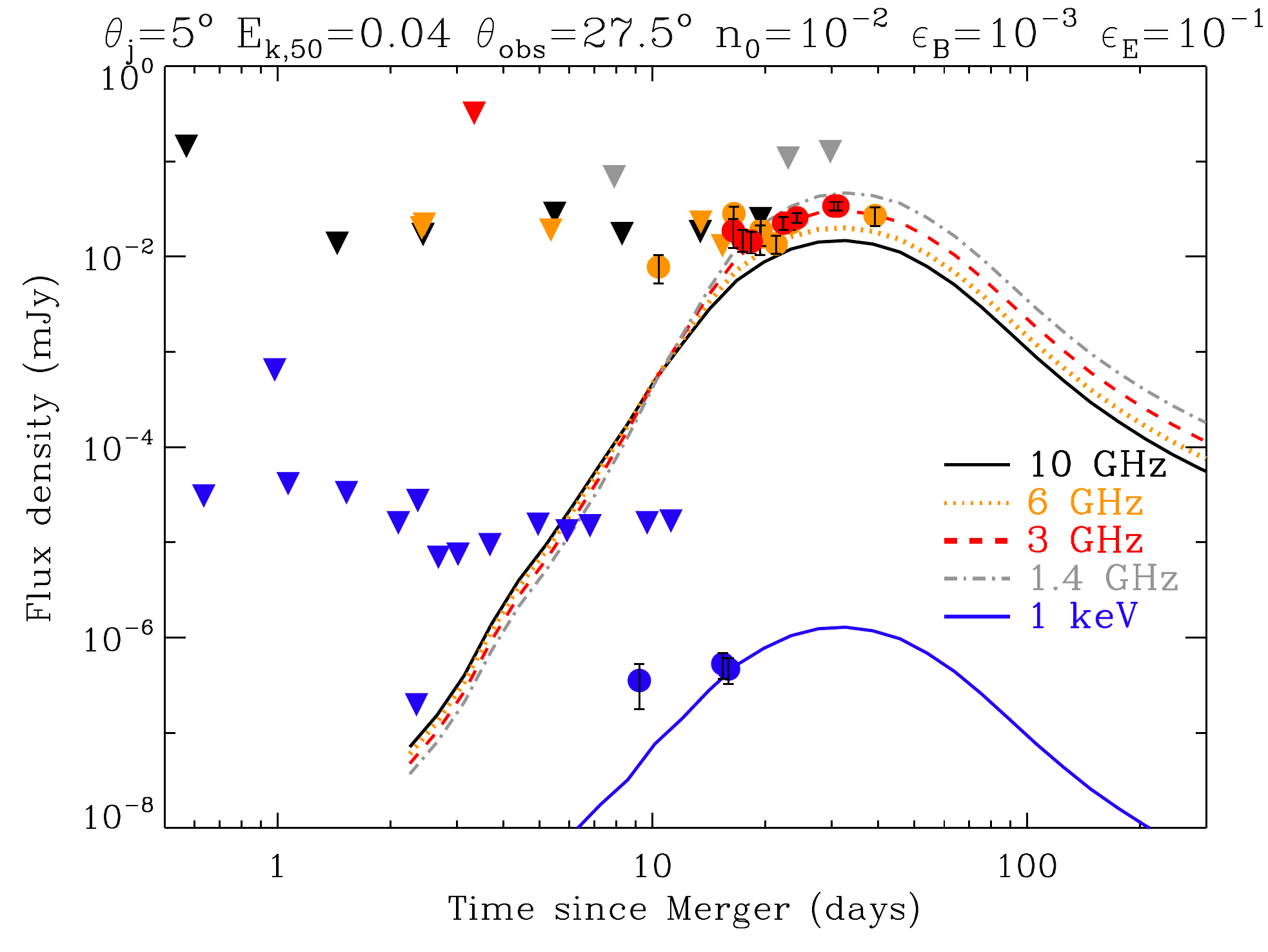}
 \includegraphics[scale=0.43]{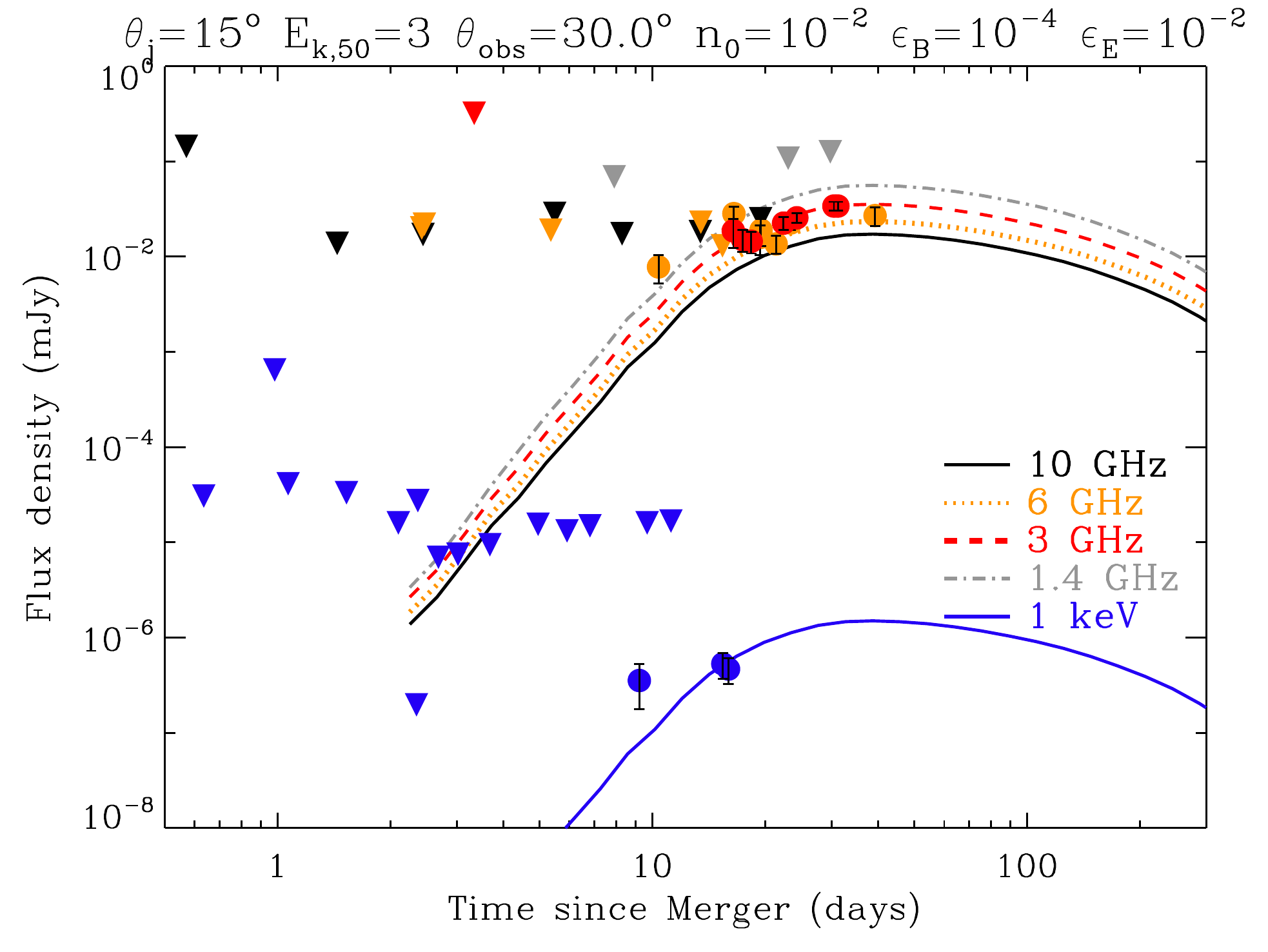}
\caption{Off-axis jet models with $\theta_j=5\degree$ (left) and $\theta_j=15\degree$ (right) that best fit the current set of X-ray (1 keV, blue) and radio observations (black, orange, red and grey for flux densities at $\sim$10 GHz, $\sim$6 GHz, $\sim$3 GHz and $\sim$1.4 GHz ,  respectively). The values of the other model parameters are listed in the plot titles. Triangles identify upper limits. These plots show the current data set and demonstrate that the emission from an off-axis relativistic uniform jet can reasonably account for the X-ray and radio observations of GW170817. Wider jets are currently favored by observations because of the milder rise and broader peak of the associated emission, as we found in A17 and M17, and as independently found by \citet{HaggardPaper}, \citet{HallinanPaper} and \citet{TrojaPaper}. Radio data at 6 GHz are displayed here for comparison, but they have not been used in our calculations (see text for details).}
\vspace*{+0.2in} 
\label{Fig:plot}
\end{figure*}

\section{Data set}
\label{Sec:data}
We collected the available X-ray and radio observations of \gw acquired at $t<40$ days since merger (Fig. \ref{Fig:plot}). This data set includes VLA observations at different frequencies ($\sim$10 GHz, 6 GHz, 3 GHz, 1.4 GHz) and X-ray observations with the Chandra X-ray Observatory (CXO). The original data sets have been published in A17, M17, \citet{HaggardPaper}, \citet{HallinanPaper}, \cite{Kim17} and \citet{TrojaPaper}. We refer to these papers for details about data acquisition and reduction. For internal consistency, we cross-calibrated the X-ray observation from \citet{TrojaPaper} using the published count-rate of 12 photons in $50$ ks of CXO observations with the spectral parameters of M17. Additionally, given the larger degree of uncertainty affecting the flux calibration of radio observations acquired at 6 GHz \citep{HallinanPaper}, we concentrate our modeling on the 1.4 GHz, 3 GHz and 10 GHz data set. The entire data set, inclusive of the 6 GHz data points, is shown in Fig. \ref{Fig:plot}.

For the measurement of $H_0$, we follow A17:H0 \nocite{LIGOH0} and use the heliocentric recessional velocity of $2995$ km/s from \cite{Kourkchi17}, as NGC 4993 can be associated as part of the group ESO-508.  Applying a bulk flow estimate from 2M++  of $-300$ km/s to the redshift in the velocity in the CMB frame, they find a final velocity of 3017 km/s.  The uncertainty for the redshift from A17:H0 is $150$ km/s.  This redshift and uncertainty is analyzed in \cite{Hjorth17} which find a lower mean heliocentric velocity of the group and a final velocity of 2922 km/s.  The magnitude of the peculiar velocity uncertainty is reconsidered below and the shift from \cite{Hjorth17} is also propagated  as a variant in the analysis.

%%%%%%%%%%%%%%%%%%%%%%%%%%%%%%%%%%%%%%%%%%
\section{Modeling of the broad-band afterglow emission}
\label{Sec:afterglow}

\begin{deluxetable}{lcc}
\tablecolumns{3}
\tablewidth{0pc}
\tablecaption{Parameters of off-axis jet simulations with {\tt BOXFIT} 
\label{tab:boxfit}}
\tablehead {
\colhead {Parameter}                &
\colhead {Range of Values}    		& 
\colhead {Grid Pace} 
}
\startdata
Jet Isotropic Energy $E_{\rm k,iso}$ (erg) & $3\times 10^{49}-3\times 10^{52}$ & 0.5 dec\\
Circum-merger density $n$ (cm$^{-3}$) & $10^{-4}-1$ & 0.5 dec\\
%Jet opening angle $\theta_{\rm j}$ (deg) & $5, 15$ \\
Observer angle $\theta_{\rm obs}$ (deg) & $(\theta_j+2.5)-90$ & 2.5$$\\
$\epsilon_B$& $10^{-4}-10^{-1}$ & 1 dec\\
$\epsilon_e$& $0.01-0.1$ & 1 dec\\
%$p$& $0.01-0.1$ & 1 dec
$p$ &$2.1-2.2$ & 0.1\\
\enddata
\tablecomments{Simulations were run at two fixed values of jet opening angles $\theta_{\rm j}=5\degree$ and $\theta_{\rm j}=15\degree$, propagating in a constant density medium. %The power-law index of the electron distribution is $p=2.1$.
 }
\label{Tab:BOXFIT}
\end{deluxetable}

We model the X-ray and radio observations of \gw with synchrotron emission from a relativistic jet pointed away from our line of sight (i.e. off-axis jet), as we did in A17 and M17 (see also \citealt{HaggardPaper,HallinanPaper,TrojaPaper}). To this aim, we run the publicly available code {\tt BOXFIT} (v2; \citealt{vanEerten10,vanEerten12}), which assumes a uniform jet with kinetic energy $E_{k}$ ploughing through a constant interstellar medium with density $n$. By varying $E_{k}$, $n$, $p$ (power-law index of the electron distribution),  $\epsilon_B$, $\epsilon_e$ (fraction of post-shock energy in magnetic fields and electrons, respectively) , and $\theta_j$ (jet opening angle), we calculate the off-axis afterglow emission as observed from different lines of sight $\theta_{obs}$, with $\theta_{obs}$ varying from $\theta_j+\delta\,\theta$ to $90\degree$ (i.e. equatorial view) with a fixed pace of $\delta\,\theta=2.5\degree$. In this paper we always refer to isotropic-equivalent luminosities, but we differentiate between isotropic-equivalent kinetic energy $E_{k,iso}$, and beaming-corrected kinetic energy of the blast wave $E_{k}$, where $E_{k}=E_{k,iso}(1-\cos{\theta_j})$.

We explore a wide portion of parameter space corresponding to $E_{k,iso}=3\times10^{49}-3\times10^{52}\,\rm{erg}$, $n=10^{-4}-1\,\rm{cm^{-3}}$, $\epsilon_B=10^{-4}-10^{-1}$, and $\epsilon_e=0.01-0.1$. Except for $\theta_{obs}$, the grid is logarithmically paced as in Table~\ref{Tab:BOXFIT}. We a priori excluded all $\theta_{obs}\leq\theta_j$ cases as clearly incompatible with the initially rising X-ray afterglow peaking at $\geq 15$~days (M17). We explored two values  of the power-law index of the electron distribution $p=2.1$ and $p=2.2$ (as expected from particle acceleration in the ultra-relativistic limit, \citealt{Sironi15}). Current X-ray and radio observations favor $p=2.1$.  We do not consider values $p>2.2$   (such as $p=2.4$, median value from short GRBs afterglows from \citealt{Fong15}) for the reasons we already discussed in M17+A17. We run each simulation for a collimated $\theta_j=5\,\degree$ jet and a jet with $\theta_j=15\,\degree$, representative of a less collimated outflow. Our choice encompasses the bulk of the distribution of estimated $\theta_j$'s for short GRBs (\citealt{Fong15} and references there in). 

To usefully limit the number of grid points in the parameter space, we preliminarily impose two observationally-driven constraints: (i) the peak time of the afterglow, calculated as $t_p=2.1\,E_{k,iso,52}^{1/3}\,n^{-1/3}\,((\theta_{obs}-\theta_j)/10\degree)^{8/3}$~days \citep{Granot02}, must lie in the range $10<t_p<300$~days; (ii) the flux density at 1~keV at $t\sim t_p$, calculated as $F_\nu(t_p)\propto \epsilon_e^{p-1}\,\epsilon_B^{(p+1)/4}\,n^{(p+1)/4}\,E_{k,iso}\,(\theta_{obs}-\theta_j)^{2(1-p)}$ (assuming X-rays lie between synchrotron peak and cooling frequency, in slow cooling regime, as expected at this epoch; e.g., \citealt{Granot02}), must match the observed flux at 15~days within a factor of $\sim30$. 
%The constraints on both observables were chosen loose enough to account for inaccuracies in the analytical vs. simulated estimates, thus retaining all the dubious cases. The points in the grid that do not match both requirements are not considered anymore, as these models overly miss basic observational constraints. {\red Raf - these last few sentences get confusing here}

The final multi-parameter posterior is estimated from the combination of a likelihood function which assigns each model $m_i$ a probability $p_i\propto\exp{(-\chi_i^2/2)}$, with $\chi_i^2$ evaluated from comparing $m_i$ with the entire broadband set of data and upper limits, along with an uninformative, scale-invariant prior on each logarithmically paced parameter. Such a prior is flat on logarithms. The final marginalized posterior on the only interesting parameter $\theta_{obs}$ is obtained by approximating the integration of the posterior over the remaining nuisance parameters space as a sum over all of the grid points. This way, the scale invariant prior is automatically encoded, as a logarithmically paced grid is equivalent to a flat distribution on logarithms.
%Median parameters for cosmological short GRBs, WF paper:
%for $\epsilon_B=0.01$: $\theta_{\rm jet}=16$ (but that assumes a lot of things like cap on opening angle; $\theta_{\rm jet}=6$ for measurements only) $n=0.04$ $E_{k,iso} = 2\times10^{51}$ which is $E_k\sim 1.6\times 10^{50}$ erg.

\begin{figure*}
\begin{center}
%\hspace*{+0.2in} 
\scalebox{1.}
{\includegraphics[width=1.0\textwidth]{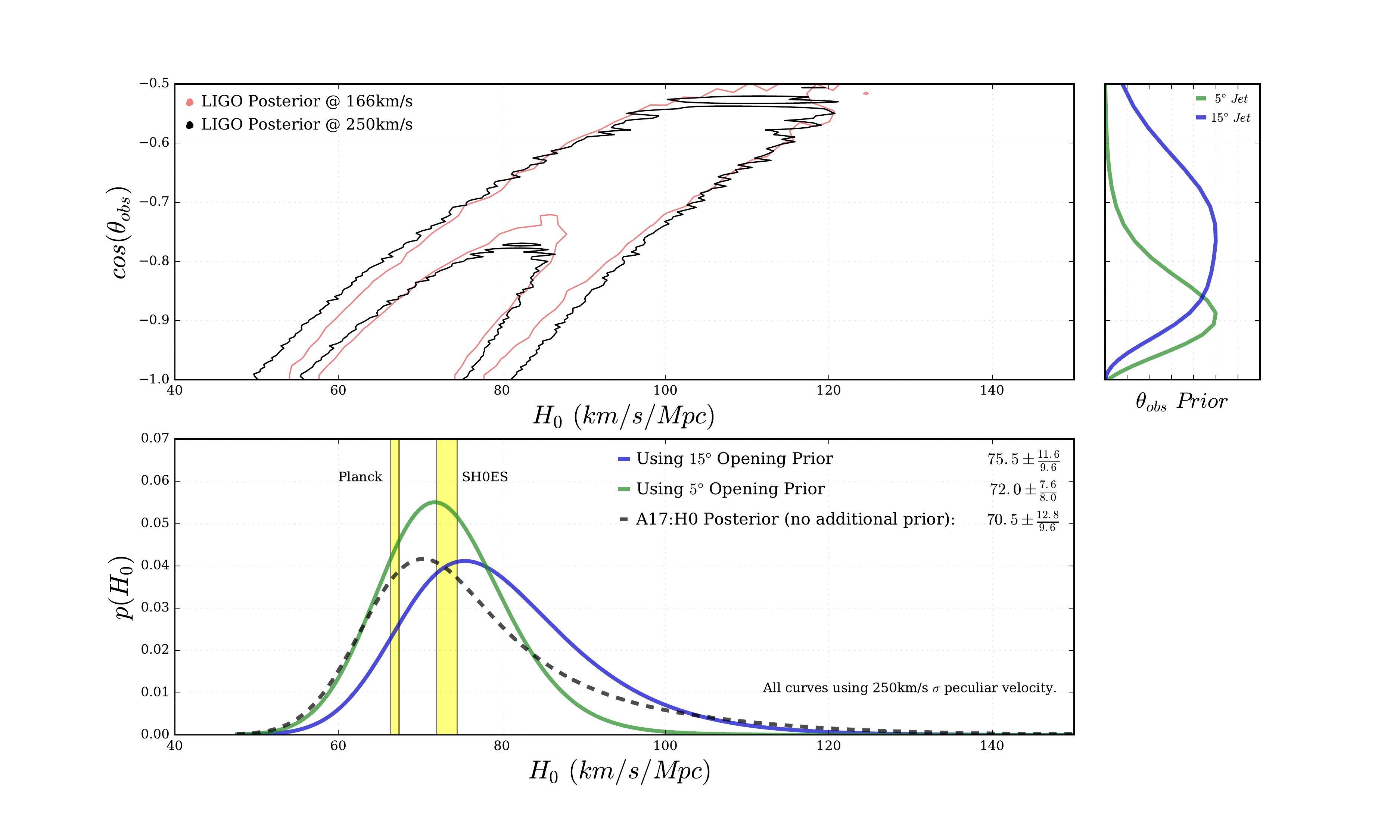}}
\caption{Constraints on the Hubble constant $H_0$ for BNS merger G298048  with and without prior on the inclination of the system. Upper: Black 1 and 2 $\sigma$ contours from LIGO data server assume 166km/s peculiar velocity uncertainty. Red 1 and 2 $\sigma$ contours noised to 250km/s peculiar velocity uncertainty. Right: Visual representation of the calculated priors on $\theta_{obs}$. Lower: Marginalized constraints on $H_0$ for our two scenarios for which inclination priors have been calculated as well as for the noised up 250km/s LIGO contour alone.}
\label{fig:constraints}
\end{center}
\end{figure*}

Overall, we confirm the results that we published in M17 and A17.
The mild temporal evolution of the X-ray and radio emission favors wider jet opening angles $\theta_j=15\degree$ (similar $\theta_j$ are invoked by \citealt{TrojaPaper,HallinanPaper} in their analysis). For $\theta_j=15\degree$, our simulations favor $n\sim10^{-4}-10^{-2}\,\rm{cm^{-3}}$, $10^{48}\,\rm{erg}<E_{k}\le 3\times 10^{50} \,\rm{erg}$ and $\theta_{obs}\sim 25\degree-50\degree$. 
%For $\theta_j=5\degree$, our modeling disfavors energetic jets with $E_{k}\ge4 \times 10^{49}\,\rm{erg}$  with $\epsilon_B\ge0.01$ and propagating in a dense medium with $n\ge0.1\,\rm{cm^{-3}}$. The allowed off-axis angles are $\theta_{obs}\sim 15\degree-40\degree$. For a wider jet of $\theta_j=15\degree$, our simulations disfavor jets with $E_{k}\ge 10^{50}\,\rm{erg}$, $\epsilon_B\ge0.01$ and $n\ge0.01\,\rm{cm^{-3}}$, and indicate $\theta_{obs}\sim 20\degree-50\degree$.  
For both jet opening angles, the measurement of $\theta_{obs}$ is degenerate with $n$ and $E_k$, with larger  $E_k/n$ fractions favoring lower $\theta_{obs}$. We show the posterior probability density function (PDF) of $\theta_{obs}$ in Fig. \ref{fig:constraints}, smoothed with a Gaussian kernel with $\sigma=7.5\degree (=3\,\delta\theta)$. 
These estimates assume a luminosity distance to NGC 4993 $d_{L}=39.5$ Mpc as listed in the NASA Extragalactic Database (NED). Varying $d_{L}$ between $36-43$ Mpc (full range of distances reported in NED, \citealt{Han92,Sakai00,Freedman01}) produces negligible impact on our final estimates. Finally, we note that the X-ray (and maybe the 6 GHz) excess at early times ($t<10$ days, Fig. \ref{Fig:plot}), might be the signature of a structured jet (i.e. a jet with energy profile that deviates from uniform and without a sharp edge, e.g. \citealt{Rossi2002,Zhang02}). Both Gaussian and power-law structured jets would be brighter than uniform jets when viewed from the same $\theta_{obs}$ before peak, while having a similar evolution at later times (see e.g. \citealt{TrojaPaper}, their Extended Data Fig. 3). 
%%%%%%%%%%%%%%%%%%%%%%%%%%%%%%%%%%%%%%%%%%
\section{Constraints on H0}
\label{Sec:H0}

%%%%%%%%%%%%%%%%%%%%%%%%%%%%%%%%%%%%%%%%%%%

%%%%%%%%%%%%%%%%%%%%%%%%%%%%%%%%%%%%%%%%%%%%%%%%%%%%%%%%%%%%%%%%%%%%%
\subsection{LIGO Data}

We use the Markov chains from A17:H0 from the 2d plane of distance and the cosine of inclination angle of the binary star system.   This is shown in Fig. 2 (top).
%%%%%%%%%%%%%%%%%%%%%%%%%%%%%%%%%%%%%%%%%%%%%%%%%%%%%%%%%%%%%%%%%%%%%
\subsection{Redshift Uncertainty}

Every galaxy responds to the pull of large-scale structure, resulting in the so-called peculiar velocity.  The observed velocity is the sum of the Hubble expansion at that redshift and the line-of-sight component of the peculiar velocity, $\mathbf{v}_{\rm obs} = \mathbf{v} + (\mathbf{v}_{\rm pec})_ {\parallel}$. To account for $(\mathbf{v}_{\rm pec})_ {\parallel}$, we adopt three alternate approaches.  Here we assume that the uncertainty in the redshift of the group by modeling the individual redshifts is sub-dominant to the peculiar velocity uncertainties.

First, we follow A17:H0 and correct the redshift by the large scale bulk flow correction from 2M++ as described in \cite{Carrick15}.  The uncertainty of this correction is $150$ km/s for halos, as estimated in N-body simulations from \cite{Carrick15}, and the uncertainty was increased in A17:H0 by $70$ km/s in quadrature due to additional correction uncertainties.  This should be considered a lower floor on the uncertainty because it assumes the ability to convert from galaxy luminosity observations to the total matter field (and thus peculiar velocity) in three dimensions --- a process that is subject to systematics because of the uncertainties in how light traces mass.

For a more conservative estimate of the peculiar velocity uncertainty, we can simply estimate the statistical variance in $H_0$ expected at $z=0.01$, without attempting to correct for it. We adopt the results from \cite{WuHuterer17} who used a large-volume cosmological N-body simulation to quantify the variance in  the local value of $H_0$ \citep{riessetal16} due to local density fluctuations and the SN sample selection.  
We use the largest volume in the public release of the Dark Sky
simulations\footnote{\href{http://darksky.slac.stanford.edu}{http://darksky.slac.stanford.edu}}
\citep{Skillman2014}, with $10240^3$ particles within a volume of $(8\hinvGpc)^3$ ($h=H_0/(100 \kmsMpc)$), and
  the mass resolution of $3.9\times10^{10} \,\hinvMsun$.   We place observers at centers of 512 sub-volumes of $(1\hinvGpc)^3$; in each
  subvolume we identify all halos in the virial mass range $[10^{12.3}, 10^{12.4}] \,\Msun$, then further identify closest-match halos to redshift $z=0.01$ in random directions on the sky. For each of these closest-match halos and in each subvolume, we measure $dH_0= (v_{\rm pec})_{\parallel}/r$.  We then calculate the variance of these measurements,
$\sigma^2((v_{\rm pec})_\parallel)$,
%   $\sigma^2(H_0)$, 
   which corresponds to the expected range due to peculiar velocity for an object at $z=0.01$ observed anywhere in the universe. The corresponding rms is
%$  \sigma(H_0)\sim  260$ km/s, 
$\sigma((v_{\rm pec})_\parallel)\approx 260$ km/s,
and it is robust with respect to simulation statistics as well as the choice of the halo mass range.

A compromise method is to apply the bulk flow corrections, but include a more conservative estimate of the peculiar velocity uncertainty as done in \cite{Scolnic17a}, which compares the dispersion in SNIa distance residuals for $z\sim0.01$ SNe and $z\sim0.05$ SNe after bulk flow corrections are applied.  The main difference in the dispersion is the impact of the peculiar velocities, which is determined to be $250$ km/s (and $270$ km/s if bulk flows were not applied). Since there is consistency between this estimate of the uncertainty and that derived based on the \cite{WuHuterer17} analysis above, we adopt $250$ km/s.  Furthermore, this estimate is consistent with the uncertainty estimated in \cite{Hjorth17} of 232 km/s. This uncertainty is $\sim8\%$ of the galaxy velocity.  

We show the inclination versus $H_0$ posterior for a peculiar velocity uncertainty of $166$ and $250$ km/s in Fig. 2 (top).  We find a small shift in the contours due to edge effects as well as the smoothing of the posterior with an additional velocity uncertainty. 
%%%%%%%%%%%%%%%%%%%%%%%%%%%%%%%%%%%%%%%%%%%%%%%%%%%%%%%%%%%%%%%%%%%%%
\subsection{Inclination Priors from EM data modeling}
Broad-band X-ray to radio observations provide an independent measurement of the inclination of the BNS system, which we quantified in Sec. \ref{Sec:afterglow} in terms of $\theta_{obs}$. NS-NS mergers are expected to launch two jets in opposite directions. $\theta_{obs}$ is the angle between the observer's line of sight and the closest of the two jets. For this system, the inclination angle is $i=180\degree -\theta_{obs}$. We re-analyze the LIGO posteriors incorporating the $\theta_{obs}$ constraints obtained from the X-ray and radio afterglow modeling to produce an updated estimate of $H_0$ (Sec. \ref{SubSec:results}). 

As a refinement, we consider additional constraints on $\theta_{obs}$ derived from modeling of the optical and near-infrared light curves and spectra of \gw (\citealt{PhilPaper,MattPaper,RyanPaper,Covino17,Diaz17,Kilpatrick17,McCully17,Pian17,Tanaka17,Tanvir17,ValentiPaper}). Explaining the observed emission requires the inclusion of an early-time ``blue'' kilonova component, consistent with lanthanide-poor ejecta, which implies $\theta_{obs} \lesssim 45\degree$ (\citealt{Kasen15,Sekiguchi16,Metzger17}). For $\theta_{obs} \gg 45\degree$ the blue kilonova emission would be obscured by the lanthanide-rich (and thus high opacity) dynamical ejecta in the equatorial plane. We note however, that this interpretation is more model-dependent and makes assumptions about the ejecta geometry. For example, this interpretation would not be necessarily true if instead the blue emission is almost always visible due to the lanthanide-poor material having a higher expansion velocity compared to the lanthanide-rich material.

%\textbf{We can infer some additional information about the inclination of the of the binary system from modeling of the optical and near-infrared light curves and spectra [CITE PHIL/MATT/RYAN]. In all cases, explaining the observed emission requires the inclusion of an early-time ``blue'' kilonova component, consistent with lanthanide-poor ejecta. In most simulations, the existence of such a component implies a viewing angle of $\theta \lesssim 45\degree$ [cite: kasen+2015, sekiguchi+2016, metzger 2017]. If the viewing angle is much larger, than the blue kilonova emission would be obscured by the lanthanide-rich (and thus high opacity) dynamical ejecta in the equatorial plane. This constraint, combined with the lower limit on $\theta$ from X-ray/Radio modeling implies a viewing angle of $20\degree \lesssim \theta \lesssim 40\degree$ [CITE MATT]. We note however, that this interpretation is model-dependent and makes assumptions about the ejecta geometry. For example, this interpretation would not be true if instead the blue emission is almost always visible due to the lanthanide-poor material having a higher expansion velocity compared to the lanthanide-rich material.}

%%%%%%%%%%%%%%%%%%%%%%%%%%%%%%%%%%%%%%%%%%%%%%%%%%%%%%%%%%%%%%%%%%%%%
\subsection{Results}
\label{SubSec:results}
Our recovered estimates of $H_0$ are summarized in Table 2.  We present a series of variations in our analysis based on the different priors from inclination and the different uncertainties on the peculiar velocity.  It is notable that in almost all of our variations, the overall best fit value of $H_0$ is higher than in A17:H0.  This is due to the higher inclination angle favored in our modeling.  The increase in $H_0$ pushes our result away from \cite{Planck16_cosmo} and towards the value measured by the SH0ES collaboration \citep{riessetal16}, though for our baseline case with a $15 \degree$ jet angle and $250$ km/s, \cite{Planck16_cosmo} is only disfavored at $\sim1\sigma$.

We also include a shifted best fit value of $H_0$ using the final velocity modeled in \cite{Hjorth17} . We performed an independent verification of their velocity estimate using the AAOmega spectrograph  (\citealt{aat1}, \citealt{aat2}), in which we obtained redshifts of 24 galaxies in the group and calculated a mean heliocentric recessional velocity  of 2953km/s. Using the peculiar velocity correction from \cite{Carrick15} we find a velocity of 2939km/s in the CMB frame. Our independent estimate agrees more with \cite{Hjorth17} than that of A17:H0 and therefore we use \cite{Hjorth17} as a variant in this analysis. In this scenario, $H_0$ decreases by $\sim3\%$, but a robust calculation requires a complex shift of the chains provide in A17:H0.   

\begin{deluxetable}{lcc}
\tablecolumns{2}
\tablewidth{0pc}
\tablecaption{Results
\label{tab:results}}
\tablehead {
\colhead {Priors}                &
\colhead {$\sigma_{vpec}$}  &
\colhead {$H_0$}
\\ 
\colhead {}    &
\colhead {(km/s)}    &
\colhead {$(\hunits)$}
}
\startdata
\vspace{0.05in} 

Baseline ($15\incangle$ Jet Width)& $250$ & \finalhbaseline\\

\vspace{0.05in}

$15\incangle$ Jet Width, v=2922km/s* & $250$ & $73.1 \pm \frac{11.3}{9.3}$\\

\vspace{0.05in} 

$15\incangle$ Jet Width& $166$& \finalhbaselow\\

\vspace{0.05in} 

$5\incangle$ Jet Width& $250$& \finalhfive\\

\vspace{0.05in} 

$5\incangle$ Jet Width& $166$& \finalhfivelow\\

\vspace{0.05in} 

$\theta_{obs}$ < $45\incangle$ (KN) & $250$ &\finalhthetalim\\

\vspace{0.05in} 

None &$250$ & \finalhligohi\\

\vspace{0.05in} 

None &$166$ & \finalhligolow\\

\vspace{0.05in}

\enddata
\tablecomments{Values reported for $H_0$ are maximum a posteriori intervals (smallest range enclosing 68\% of the posterior). * Denotes heliocentric velocity obtained from \cite{Hjorth17} of 2922km/s. All other values reported use A17:H0 velocity estimate of 3017km/s.}
\label{Tab:results}
\end{deluxetable}

%%%%%%%%%%%%%%%%%%%%%%%%%%%%%%%%%%%%%%%%%%%
\section{Summary and Conclusions}
\label{Sec:Conc}

%\newcommand{\finalhbaseline}{$75.5 \pm \frac{11.9}{8.2}~$}
%\newcommand{\finalhligohi}{$70.5 \pm \frac{13.1}{7.4}~$}
%\newcommand{\finalhligolow}{$69.7 \pm \frac{12.4}{5.4}~$}
%\newcommand{\finalhfive}{$71.8 \pm \frac{8.7}{7.2}~$}
%\newcommand{\finalhthetarange}{$73.7 \pm \frac{8.6}{7.4}~$}
%\newcommand{\finalhthetalim}{$74.5 \pm \frac{12.6}{7.7}~$}

%\finalhfive
%\finalhbaseline

In this study, we combined constraints on the inclination from modeling of the electromagnetic observations  of \gw with the correlated constraints between distance and inclination as presented in A17:H0.  While our inclination constraints from modeling of EM data are model dependent, our best fit case assuming the favored $15 \degree$ jet angle yields $H_0=$\finalhbaseline~\hunits.  If we assume $5 \degree$ jet angle, we measure $H_0=$\finalhfive~\hunits.  Both of these measurements are shifted towards higher $H_0$ values from $H_0=$\finalhabbott~\hunits\ in A17:H0.  Our uncertainty on $H_0$ for the $15 \degree$ case is not significantly reduced when applying the inclination constraints, partly  due to our increase in the peculiar velocity uncertainty from A17:H0.  Still, it can be noted our recovered $H_0$ value marginally favors the \cite{riessetal16} over the \cite{Planck16_cosmo} by $\sim1\sigma$.

We find that the leverage from the independent inclination measurement reduces the uncertainty on $H_0$ from the distance alone from $\sim14.5$ km/s to $\sim 11.7$ km/s, assuming a symmetric uncertainty\footnote{larger than uncertainty from MAP parameterization because MAP minimizes 68\% region}.  The measurement uncertainty from \cite{riessetal16} is 1.7 km/s.  To reach that level of precision using GW measurements, with similar X-ray/radio constraints, 47 events like this one would be needed, but without X-ray/radio constraints, 72 events are needed.  Given expected rates of $1$ KN per year discovered by LIGO at current sensitivity, the importance of X-ray/radio measurements is obvious.

This is especially true if like as expected for this specific case of \gw, the constraints on the inclination are significantly improved with observations at $t \gtrsim 100$ days since merger, when \gw will be observable again in the X-rays. %This event was too close to the Sun. 
If GW events are similar to \gw, it will be possible to detect future GW events at X-rays and radio wavelengths out to a distance of $\sim100$ Mpc\footnote{It will be possible to detect similar systems at $d>100$ Mpc only if the jet is closer to our line of sight, so likely won't be able to follow the large numbers of KN detections predicted in \cite{Scolnic17b}}.  Since the inclination constraint from X-ray and radio data gets worse at further distances, but the impact of uncertainties in the peculiar velocities on $H_0$ is inversely proportional to distance, the overall $H_0$ uncertainty of future events is likely to be similar to our calculated uncertainty for \gw.  For a very competitive measurement of $H_0$ from GW events, both the sensitivity of GW detectors as well as X-ray and radio detectors must increase.

%For future detections, it is likely that the KNe are fainter than the one discovered, so radio and x-ray observations will require resources like {\red Raf}. 

%%%%%%%%%%%%%%%%%%%%%%%%%%%%%%%%%%%%%%%%%%%
\bigskip
C.G. acknowledges University of Ferrara for use of the local HPC facility co-funded by the ``Large-Scale Facilities 2010'' project (grant 7746/2011). This research was supported in part through the computational resources and staff contributions provided for the Quest high performance computing facility at Northwestern University which is jointly supported by the Office of the Provost, the Office for Research, and Northwestern University Information Technology. We gratefully acknowledge Piero Rosati for granting us usage of proprietary HPC facility. The Berger Time-Domain Group at Harvard is supported in part by the NSF through grants AST-1411763 and AST-1714498, and by NASA through grants NNX15AE50G and NNX16AC22G. DAB is supported by NSF award PHY-1707954. Development of the Boxfit code was supported in part by NASA through grant NNX10AF62G issued through the Astrophysics Theory Program and by the NSF through grant AST-1009863. Simulations for BOXFITv2 have been carried out in part on the computing facilities of the Computational Center for Particle and Astrophysics of the research cooperation ``Excellence Cluster Universe'' in Garching, Germany. D.S. is supported by NASA through
Hubble Fellowship grant HST-HF2-51383.001 awarded by the Space Telescope Science Institute, which is operated by the Association of Universities for Research in Astronomy, Inc., for NASA, under contract NAS 5-26555.
We gratefully acknowledge the director of the Anglo-Australian Telescope for their discretionary time on the AAOmega Spectrograph. Based in part on data acquired through the Australian Astronomical Observatory. We acknowledge the traditional owners of the land on which the AAT stands, the Gamilaraay people, and pay our respects to elders past and present.

%%%%%%%%%%%%%%%%%%%%%%%%%%%%%%%%%%%%%%%%%%%
%\bibliographystyle{apj}
%\bibliography{margutti}

\end{document}